\documentclass[runningheads]{llncs}
\usepackage[T1]{fontenc}
\usepackage{graphicx}
\usepackage{booktabs}
\usepackage{color} 
\usepackage{verbatim}
\usepackage{CJK}
\usepackage{amsmath}
\usepackage{amssymb}
\usepackage{algorithm} 
\usepackage{algpseudocode}
\usepackage[colorlinks,linkcolor=blue,anchorcolor=blue,citecolor=blue,urlcolor=blue]{hyperref}

\usepackage{graphicx,amsmath,multirow,bm,color,xr}
\usepackage{graphicx,amstext,amsfonts,bm}
\usepackage[misc]{ifsym}
\usepackage[colorlinks,urlcolor=blue, citecolor=blue, linkcolor=blue]{hyperref}
\makeatletter \def\thanks#1{\protected@xdef\@thanks{\@thanks \protect\footnotetext{#1}}} \makeatother

\begin{document}
%
% \maketitle
\title{A Novel Unified Conditional Score-based Generative Framework for Multi-modal Medical Image Completion}
\author{Xiangxi Meng\inst{1, 2, 3}\thanks{X. Meng and Y. Gu - Equal contribution.}\and
Yuning Gu\inst{1}\and
Yongsheng Pan\inst{1}\and
Nizhuan Wang\inst{1}\and
Peng Xue\inst{1}\and
Mengkang Lu\inst{4}\and
Xuming He\inst{2}\and
Yiqiang Zhan\inst{3}\and
Dinggang Shen\inst{1,3}$^{(\textrm{\Letter})}$}

\authorrunning{X. Meng et al.}

\institute{School of Biomedical Engineering, ShanghaiTech University, Shanghai, China
\email{dgshen@shanghaitech.edu.cn} \\
\and
School of Information Science and Technology, ShanghaiTech University, Shanghai, China \and
Shanghai United Imaging Intelligence Co., Ltd., Shanghai, China \and
National Engineering Laboratory for Integrated Aero-Space-Ground-Ocean Big Data Application
Technology, School of Computer Science and Engineering, Northwestern Polytechnical University,
Xi’an, Shaanxi, China}

% \email{{\{abc,lncs\}@uni-heidelberg.de}}
%
\maketitle              % typeset the header of the contribution
\begin{abstract}
Multi-modal medical image completion has been extensively applied to alleviate the missing modality issue in a wealth of multi-modal diagnostic tasks. However, for most existing synthesis methods, their inferences of missing modalities can collapse into a deterministic mapping from the available ones, ignoring the uncertainties inherent in the cross-modal relationships. Here, we propose the Unified Multi-Modal Conditional Score-based Generative Model (UMM-CSGM) to take advantage of Score-based Generative Model (SGM) in modeling and stochastically sampling a target probability distribution, and further extend SGM to cross-modal conditional synthesis for various missing-modality configurations in a unified framework. Specifically, UMM-CSGM employs a novel multi-in multi-out Conditional Score Network (mm-CSN) to learn a comprehensive set of cross-modal conditional distributions via conditional diffusion and reverse generation in the complete modality space. In this way, the generation process can be accurately conditioned by all available information, and can fit all possible configurations of missing modalities in a single network. Experiments on BraTS19 dataset show that the UMM-CSGM can more reliably synthesize the heterogeneous enhancement and irregular area in tumor-induced lesions for any missing modalities.
\end{abstract}
\section{Introduction}

Multi-modal medical imaging allows aggregation of complementary modality-specific information for comprehensive evaluation of the pathological status. However, not all desired imaging modalities can be acquired due to practical limitations including long scan time \cite{22}, image corruption \cite{27}, etc. This missing modality issue hinders the usage of multi-modal data for diagnostic purposes \cite{4,12}. A promising solution, which has been demonstrated to alleviate miss-modality issue in multi-modal diagnostic tasks (e.g., multi-modal brain tumor segmentation \cite{1}), is to synthesize missing modalities from available ones by exploiting the cross-modal relationships as prior knowledge \cite{23,6,7,9}. 

Similar to many inverse problems \cite{5,16}, the cross-modal relationships inherently contain uncertainties, since only redundant modalities can be completely synthesized from the available ones. Such uncertainties make it necessary to capture the cross-modal conditional probability distribution, so that the confidence in synthesized images can be known. Further, all available modalities should be leveraged to accurately condition the distribution of the missing ones for efficient image synthesis. Meanwhile, to address the real-world situations with variable missing modalities, a unified framework is desired to learn a comprehensive set of conditional distributions.

Extensive efforts have been devoted to exploiting deep-learning methods, especially those based on conditional generative adversarial network (cGAN) \cite{9,8,28,15,26}, to perform cross-modal image completion. The cGAN based methods train a generator to translate available-modality images to missing-modality images by following the cross-modal conditional distribution implicitly learned by a discriminator. However, in practice, the cross-modal relationships learned by cGAN-based method (as shown in Pix2Pix \cite{31}) can degrade into a deterministic mapping that outputs the average of the probable images weighted by their conditional probability, yielding suboptimal results.

Recently, the score-based generative model (SGM) \cite{7,20,18,19,10,3} achieves great success in image generation, owing to their capability in capturing and effectively sampling the target distribution using the stochastic diffusion (i.e., Markovian transition to noise distribution) and reverse generation (i.e., denoising) strategy \cite{3}. Owing to its probabilistic nature, SGM has been shown to outperform even the best-in-class GAN-based methods in generating images with improved fidelity and diversity \cite{6,13,25}. However, the original SGM performs unconditional intra-modality image generation, and thus most existing conditional synthesis variants of SGM are limited to a predefined single-source single-target configuration. Therefore, these methods cannot allow cross-modal image completion in a unified framework.

Here, we present a novel generative framework called Unified Multi-Modal Conditional Score-based Generative Model (UMM-CSGM), which for the first time, to our best knowledge, generalizes SGM for unified cross-modal conditional image completion. Our contributions are: 1) We propose to learn cross-modal image synthesis via conditional diffusion and score-based reverse generation defined in the complete multi-modality space. 
The conditional generation of any possible missing modalities are learned by reversing the diffusion process applied to the modalities to be synthesized, given all remaining modalities stay unchanged as conditions. 2) We develop a novel multi-in multi-out Conditional Score Network (mm-CSN) to reverse the diffusion process in the complete modality space for unified conditional generation.
The mm-CSN enables using a single network to gradually refine the synthesized images of any missing modalities from noise distribution, with the entire synthesis process accurately conditioned by all available modalities. 3) Our experimental results on the BraTS19 dataset show that UMM-CSGM can better model the cross-modal relationships to synthesize missing-modality images with higher similarity to the ground truth, as compared to the existing state-of-the-art (SOTA) generative methods.

\begin{figure}  
\centering  
\includegraphics[width=1\textwidth]{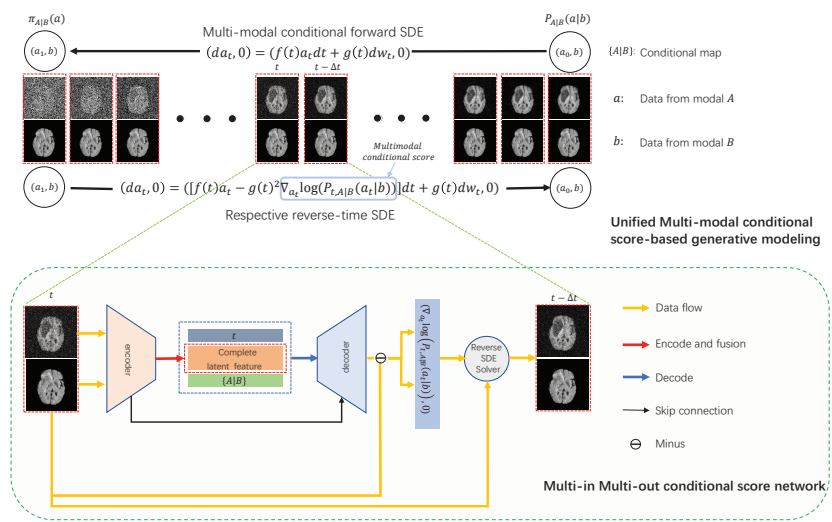}
\caption{Schematics of Unified Multi-Modal Conditional Score-based Generative Model (UMM-CSGM). The upper panel shows the cross-modal conditional diffusion and reverse generation process. The lower panel shows the structure of multi-in multi-out Conditional Score Network (mm-CSN).}
\label{fig1}
\end{figure} 

\section{Method}
Since multi-modal images compose complementary views of the same subject, cross-modal relationships formulated by deterministic mappings generally do not hold. In this section, we present the UMM-CSGM framework which can handle the uncertainties inherent in the cross-modal relationships by capturing various cross-modal conditional distributions using a single multi-in multi-out conditional score network.

\subsection{Unified multi-modal conditional score-based generative model}
\subsubsection{Naive diffusion and score-based reverse generation.}

SGM progressively generates data from noise via learning to reverse a diffusion process that gradually diffuse the data into noise.
The diffusion process can be formulated as a forward Stochastic Differential Equations (SDEs)
\begin{equation}
    \mathrm{d}\mathbf{x}_{t}=f(t)\mathbf{x}_{t}\mathrm{d}t+g(t)\mathrm{d}\mathbf{w}_{t}, t \in \left[0, 1\right].
    \label{Eq1}
\end{equation}
Here, $f$ is a linear drift coefficient, $g$ is a scalar diffusion coefficient, and $\mathbf{w}$ is a standard Wiener process. These variables are designed to transfer data $\mathbf{x}_0 \sim P_0(\mathbf{x})$ into pure noise $\mathbf{x}_1 \sim \pi(\mathbf{x})$, where $\pi(\mathbf{x})$ is a predefined Gaussian distribution.

Accordingly, the reverse generation process gradually recovers the data distribution from noise ($\pi(\mathbf{x}) \rightarrow P_0(\mathbf{x})$) by applying reverse SDE:
\begin{equation}
    \mathrm{d}\mathbf{x}_{t}=\left[f(t)\mathbf{x}_{t}-g(t)^{2}\nabla_{\mathbf{x}_{t}}log(P_t(\mathbf{x}_{t}))\right]\mathrm{d}t+g(t)\mathrm{d}\mathbf{\bar{w}}_{t},\label{Eq2}
\end{equation}
where the score $\nabla_{\mathbf{x}_t}log(P_t(\mathbf{x}_t))$ for each $t$ can be estimated using the denoising score matching method \cite{24}, given the clean samples in $P_0(\mathbf{x})$ and noisy samples in $P_t(\mathbf{x}_t)$
generated in the forward SDE. More details can be found in \cite{22,15}.

\subsubsection{Unified multi-modal conditional diffusion and reverse generation.} 
The schematics of UMM-CSGM are illustrated in the Fig.~\ref{fig1} (upper panel). Without loss of generality, let's assume that the imaging modalities $\mathcal{C}$ can be grouped into two sets: 1) set $\mathcal{A}$ composed of modalities to be synthesized, and 2) set $\mathcal{B} = \mathcal{C} -  \mathcal{A}$ involving conditional modalities. The modalities in sets $\mathcal{A}$ and $\mathcal{B}$ can be randomly partitioned. Let $(\mathbf{a}, \mathbf{b})$ represent paired multi-modal images from the same subject in the joint cross-modal space $\mathcal{A}\times \mathcal{B}$, and $\mathcal{S}:\{\left(\mathbf{a}, \mathbf{b} \right): \left(\mathbf{a}, \mathbf{b}\right) \in \mathcal{A}\times \mathcal{B}\}$ denote the aggregation of paired images from the two modality sets.

In conditional diffusion process, we sequentially diffuse images $\mathbf{a}$ with $\mathbf{a}_0 \sim P_{0,\mathcal{A}|\mathbf{B}}(\mathbf{a}|\mathbf{b})$ into $\mathbf{a}_1 \sim \pi_{\mathcal{A}|\mathbf{B}}(\mathbf{a})$ by slowly increasing the noise, while leaving images $\mathbf{b}$ unchanged. Then, we learn to reverse the diffusion process by gradually refining $\mathbf{a}$ under the guidance of $\mathbf{b}$ (i.e., $\pi_{\mathcal{A}|\mathcal{B}}(\mathbf{a}) \rightarrow P_{0,\mathcal{A}|\mathcal{B}}(\mathbf{a}|\mathbf{b})$).  

The proposed multi-modal conditional diffusion process can be formulated as a multi-modal conditional forward SDE as shown in Eq.\ref{Eq3}:
\begin{equation}
    % \mathrm{d}\mathbf{b}=f((\mathbf{a}, \mathbf{b}), \mathbb{t})\mathrm{d}\mathbb{t}+g(\mathbb{t})\mathbf{w}
    \mathrm{d} (\mathbf{x}_{\{{\mathcal{A}|\mathcal{B}},t\}})=(\mathrm{d}\mathbf{a}_{t},\mathrm{d}\mathbf{b}_{t} \equiv 0)=(f(t)\mathbf{a}_{t}\mathrm{d}t+g(t)\mathrm{d}\mathbf{w}_{t}, 0), t \in \left[0, 1\right].
    \label{Eq3}
\end{equation}
Accordingly, the reverse-time process of the proposed diffusion process can be formulated as a multi-modal conditional reverse SDE in Eq.\ref{Eq4}:
\begin{equation}
    % \mathrm{d}\mathbf{b}=f((\mathbf{a}, \mathbf{b}), \mathbb{t})\mathrm{d}\mathbb{t}+g(\mathbb{t})\mathbf{w}
    \begin{split}
        \mathrm{d} (\mathbf{x}_{\{{\mathcal{A}|\mathcal{B}},t\}})&=(\mathrm{d}\mathbf{a}_{t}, \mathrm{d}\mathbf{b}_{t} \equiv 0) \\
        &=(\left[f(t)\mathbf{a}_{t}-g(t)^{2}\nabla_{\mathbf{a}_{t}}log(P_{t,\mathcal{A}|\mathcal{B}}(\mathbf{a}_{t}| \mathbf{b})\right]\mathrm{d}t+g(t)\mathrm{d}\mathbf{\bar{w}}_{t}, 0).
    \end{split}
        \label{Eq4}
\end{equation}
The cross-modal conditional score function  
$\nabla_{\mathbf{a}_{t}}log(P_{t,\mathcal{A}|\mathcal{B}}(\mathbf{a}_{t}|\mathbf{b}))$ can be estimated via denoising score matching \cite{24} given the clean images $(\mathbf{a}, \mathbf{b})$ and partially diffused images $(\mathbf{a}_t, \mathbf{b})$.

Note that conditional diffusion and score-based reverse generation is set in the complete multi-modality space. Therefore, UMM-CSGM can become a unified framework to reverse the diffusion process of any modalities under the guidance of all remaining modalities, with a single network trained to learn various cross-modal conditional score function.

\subsection{Multi-in multi-out conditional score-based generative network}
\subsubsection{Network design.}
As shown in the lower panel of Fig.~\ref{fig1}, mm-CSN adopts a multi-in multi-out U-Net structure \cite{28} to learn the multi-modal conditional score function at each time step $t$. Specifically, the paired multi-modal images $(\mathbf{a}_t, \mathbf{b})$ in the complete modality space are fed into the encoder. The latent features extracted by the encoder, together with current time step $t$ and a code notifying the conditional configuration $\{\mathbf{A}|\mathbf{B}\}$, are sent to the decoder. The decoder then yields the time-dependent multi-modal conditional score function $(\nabla_{\mathbf{a}_{t}}log(P_{t,\mathcal{A}|\mathcal{B}}(\mathbf{a}_{t}|\mathbf{b})),0)$. Functionality of the mm-CSN is formulated in Eq.\ref{Eq5}:
\begin{equation}
    \mathbf{NN}_{\theta}((\mathbf{a}_{t}, \mathbf{b}),t, \mathcal{A}|\mathcal{B})=(\mathbf{a}_{t}, \mathbf{b})+( \nabla_{\mathbf{a}_{t}}log(P_{t,\mathcal{A}|\mathcal{B}}(\mathbf{a}_{t}|\mathbf{b})), 0)
    \label{Eq5}
\end{equation}
where $\mathbf{NN}_{\theta}$ represents the mm-CSN parameterized by $\theta$. Switching the conditional modalities and time step $t$ can therefore yield various multi-modal conditional scores using a single network. Such compact and light-weight network design can also enhance the effectiveness in training and inference. 

\subsubsection{Learning the multi-modal conditional score.}
The training objective for the conditional configuration $\{\mathcal{A}|\mathcal{B}\}$ at time $t$ is formulated based on denoising score matching method \cite{24}:
\begin{equation}
% \begin{align}
    \begin{split}
        \mathcal{L}_{\mathbb{DSM},\mathcal{A}|\mathcal{B}}(\theta, t)&= 
        \frac{1}{2}\mathbb{E}_{\substack{\mathbf{b} \sim P_{\mathcal{B}}(\mathbf{b})\\ \mathbf{a} \sim P_{\mathcal{A}|\mathcal{B}}(\mathbf{a}|\mathbf{b})\\ 
        \mathbf{a}_{t} \sim P_{0t}(\mathbf{a}_{t}|\mathbf{a}, \mathbf{b})}}
        \Vert(\mathbf{a}_{t}+ S(\mathbf{a}_{t}), \mathbf{b})-\mathbf{NN}_{\theta}((\mathbf{a}_{t}, \mathbf{b}), t, \mathcal{A}|\mathbf{B})\Vert_{2}^{2} \\
        s.t. \ \  S(\mathbf{a}_{t})&=\nabla_{\mathbf{a}_{t}}log(P_{0t}(\mathbf{a}_{t}|\mathbf{a}, \mathbf{b))} \label{Eq6}
    \end{split}
%\nabla_{\mathbf{a}_{t}}log(P_{0t}(\mathbf{a}_{t}|\mathbf{a}, \mathbf{b))}
% \end{align}
\end{equation}
The loss $\mathcal{L}_{\mathbb{DSM},\mathcal{A}|\mathcal{B}}(\theta, t)$ ensures that the network $\mathbf{NN}_{\theta}$ can learn the time-dependent conditional score function $( \nabla_{\mathbf{a}_{t}}log(P_{t,\mathcal{A}|\mathcal{B}}(\mathbf{a}_{t}|\mathbf{b})), 0)$  described in Eq.\ref{Eq4}. 

Considering all possible possible configurations of set $\mathcal{A}$ and $\mathcal{B} = \mathcal{C} -  \mathcal{A}$ in the time interval $t \sim U(0,1)$, we can define the overall training objective in Eq.\ref{Eq7}:
\begin{equation}
    \mathcal{L}_{\mathbb{DSM}}(\theta)=\frac{1}{N_\mathcal{A}}\sum\limits_{\substack{\mathcal{A} \subset \mathcal{C} \\ \mathcal{B}=\mathcal{C}-\mathcal{A}}}\mathbb{E}_{t \sim U(0, 1)}\left[(\mathcal{L}_{\mathbb{DSM},\mathcal{A}|\mathcal{B}}(\theta, t)\right],
    \label{Eq7}
\end {equation} 
where $N_\mathcal{A}$ is the number of modality configurations in set $\mathcal{A}$. With $\mathbf{NN}_{\theta}$ trained on all modality configurations, unified image generation using UMM-CSGM framework can be realized as described in Algorithm~\ref{alg1}.
\begin{algorithm}[htp]
\caption{Multi-modal conditional score-based image generation}     
\label{alg1}       
\hspace*{0.02in} {\bf Input:} 
% A bag of instance embeddings $H\in \mathbb{R}^{N\times D}$
{Conditional images $\mathbf{b}$, conditional configuration $\{\mathcal{A}|\mathcal{B}\}$, total time step $T$}
\begin{algorithmic}[] 
\State $\mathbf{a}_{1} \sim \pi_{\mathcal{A}|\mathcal{B}}(\mathbf{a}), \Delta t= \frac{1}{T}$
% \State $H_{0}$\leftarrow $\mathrm{Concat}\left(H_{p},H\right)$, where $H_{p}\in \mathbb{R}^{1\times D}$
\For {$t>=0$}
    \State $\mathbf{a}_{t-\Delta t} \leftarrow \mathbf{a}_{t}-f(t)\mathbf{a}_{t}\Delta t+g(t)^{2}\left[\mathbf{NN}_{\theta}((\mathbf{a}_{t}, \mathbf{b}), t, \mathcal{A}|\mathcal{B})-(\mathbf{a}_t, \mathbf{b})\right]\Delta t$
    \State $n \sim \mathcal{N}(0, \mathbf{I})$
    \State $\mathbf{a}_{t-\Delta t} \leftarrow \mathbf{a}_{t-\Delta t} + g(t)n\sqrt{\Delta t},$ $t \leftarrow t-\Delta t$
\EndFor
\end{algorithmic} 
\hspace*{0.02in} {\bf Output:} 
$\mathbf{a}_{0} \sim P_{\mathcal{A}|\mathbf{B}}(\mathbf{a}|\mathbf{b})$
\end{algorithm}

\section{Experiments}
\subsection{Experimental setup}
\subsubsection{Materials and evaluation metrics.}
We perform multi-modal brain MRI image completion on the BraTS19 dataset. Each subject has four MRI modalities, i.e., Flair, T1, T1c, and T2. We randomly divide the data set into training set, validation set, and test set with a ratio of 20/1/8. The generation tasks are set as missing any one of the modalities, while the remaining three modalities are available. The multi-modal conditional diffusion and reverse generation process is performed with a total time step $T=1000$. Three metrics are used to evaluate the quality of the synthesized images. PSNR is calculated by $20 log(MAX_I)/\sqrt{MSE}$, where $MAX_I$ is the highest possible intensity value in the image and $MSE$ is the mean-squared-error between two images. SSIM computes structural similarity between two images \cite{30}. MAE computes the mean absolute error between two images. Higher PSNR, SSIM, and lower MAE indicate a higher quality of the generated image with respect to the ground truth.

\subsection{Comparison with state-of-the-art methods}
We compare our proposed UMM-CSGM method to several representative SOTA methods developed for multi-modal medical image completion, including cross-modal mapping methods and common-space methods. Briefly, cross-modal mapping methods, e.g., CocaGAN \cite{6}, Hi-Net \cite{7}, and MM-GAN \cite{9}, usually learn a deterministic mapping from source to target modality. Common space methods, e.g., UcDGAN \cite{8} and MM-SYN \cite{29}, typically build a latent space where different modalities of the same subject can be represented by a common feature, and leverage this latent common space as the hub for cross-modal translation.

\begin{table}[!t]
\centering
\caption{Evaluation metrics (mean ± standard deviation) of the synthesized \textbf{Flair}/ \textbf{T1}/ \textbf{T1c}/ \textbf{T2} images on the BraTS19 dataset using different conditional generative methods. All remaining modalities are used as conditions. Bold indicates the best result.}
\begin{tabular}{l|ccc|ccc} 
% \footnotesize
% \begin{tabular}{cccc} 
   \toprule
   & & \textbf{Flair}  &  & & \textbf{T1}  &  \\
Method  & PSNR $\uparrow$ & SSIM $\%$ $\uparrow$ & MAE $\downarrow$  & PSNR $\uparrow$ & SSIM $\%$ $\uparrow$ & MAE $\downarrow$ \\
\hline
Ours   &   \textbf{25.73}$\pm$\textbf{1.94}   &   \textbf{98.13}$\pm$\textbf{0.60}   &   \textbf{5.32}$\pm$\textbf{1.29}  &   \textbf{29.92}$\pm$\textbf{1.47}   &   \textbf{98.92}$\pm$\textbf{0.24}   &   \textbf{3.41}$\pm$\textbf{0.59}    \\
MM-GAN   &   23.98$\pm$2.76   &   97.74$\pm$0.78   &   6.61$\pm$1.95  &   28.01$\pm$1.63   &   98.64$\pm$0.36   &   4.33$\pm$0.82    \\
UcDGAN   &   23.70$\pm$2.69   &   97.72$\pm$0.77   &   6.74$\pm$1.96   &   27.44$\pm$1.62   &   98.56$\pm$0.39   &   4.54$\pm$0.91   \\
Hi-Net   &   24.02$\pm$2.81   &   97.88$\pm$0.85   &   6.56$\pm$1.99   &   28.44$\pm$1.58   &   98.68$\pm$0.34   &   4.10$\pm$0.79   \\
MM-SYN   &   21.66$\pm$1.19   &   96.58$\pm$0.63   &   8.41$\pm$1.33  &   23.12$\pm$1.40   &   96.82$\pm$0.58   &   7.20$\pm$1.36    \\
CocaGAN   &   23.14$\pm$1.31   &   97.41$\pm$0.69   &   7.64$\pm$1.48  &   26.28$\pm$1.73   &   98.16$\pm$0.54   &   5.20$\pm$1.11   \\
\hline\hline
   & & \textbf{T1c} &  & & \textbf{T2}  &  \\
Method  & PSNR $\uparrow$ & SSIM $\%$ $\uparrow$ & MAE $\downarrow$  & PSNR $\uparrow$ & SSIM $\%$ $\uparrow$ & MAE $\downarrow$ \\
\hline
Ours   &   \textbf{27.81}$\pm$\textbf{1.65}   &   \textbf{98.57}$\pm$\textbf{0.48}   &   \textbf{4.08}$\pm$\textbf{1.16} & \textbf{26.41}$\pm$\textbf{1.72}   &   \textbf{98.52}$\pm$\textbf{0.73}   &   \textbf{4.83}$\pm$\textbf{1.30}  \\
MM-GAN   &   26.24$\pm$1.74   &   98.31$\pm$0.54   &   5.02$\pm$1.22  &   25.09$\pm$1.65   &   98.19$\pm$0.80   &   5.81$\pm$1.50    \\
UcDGAN   &   25.99$\pm$1.73   &   98.27$\pm$0.56   &   5.10$\pm$1.28  &   24.49$\pm$1.47   &   98.11$\pm$0.76   &   6.14$\pm$1.44    \\
Hi-Net   &   26.31$\pm$1.76   &   98.33$\pm$0.56   &   4.98$\pm$1.32 &   24.73$\pm$1.77   &   98.12$\pm$0.89   &   5.90$\pm$1.61    \\
MM-SYN   &   18.54$\pm$1.45   &   92.65$\pm$0.17   &   12.51$\pm$2.05  &   18.26$\pm$1.28   &   94.31$\pm$0.12   &   12.14$\pm$1.99   \\
CocaGAN   &   25.12$\pm$1.80   &   97.82$\pm$ 0.69   &   5.66$\pm$1.50  &   18.93$\pm$1.54   &   95.32$\pm$1.62   &   12.62$\pm$2.75   \\
 \bottomrule
\end{tabular}
%}
\label{Tab1}
\end{table}

\begin{figure}[!t] 
\centering  
\includegraphics[width=1\textwidth]{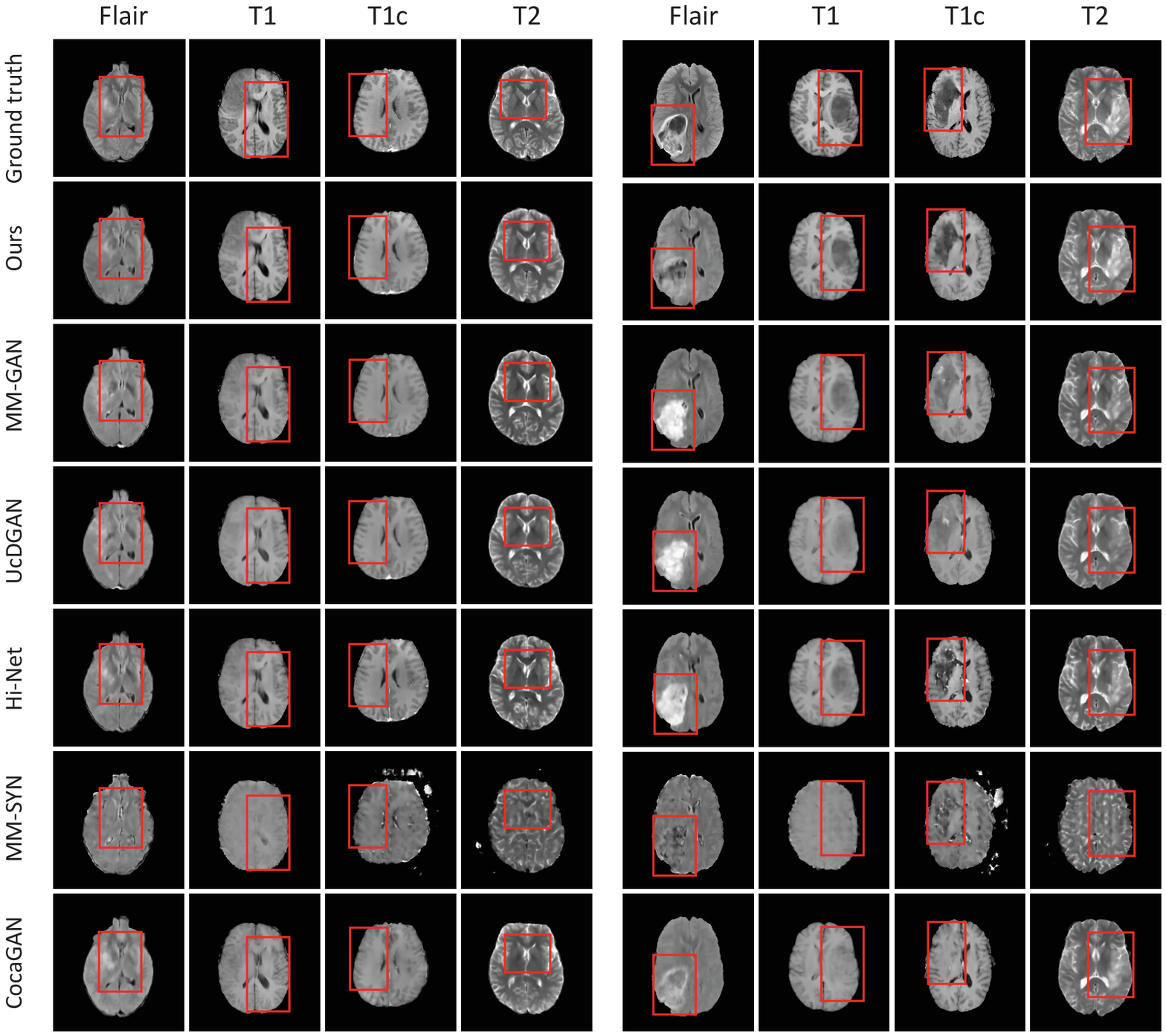}
\caption{Visual comparison of synthesized MRI image. The left panel shows representative synthesis results with relatively benign tumors, while the right panels shows results with more malignant tumors. The columns of each panel from left to right show Flair, T1, T1c, and T2 images. The rows from bottom to up denote the synthesis results of different methods, i.e., CocaGAN, MM-SYN, Hi-Net, UcDGAN, MM-GAN, Ours, and the ground truth (the top row).}  
\label{fig2}  
\end{figure}

\subsubsection{Quantitative results.}
We report the evaluation metrics of competing generative methods (i.e., CocaGAN, MM-SYN, Hi-Net, UcDGAN, MM-GAN, UMM-CSGM) for various missing-modality configurations including missing Flair, missing T1, missing T1c, and missing T2 in Table~\ref{Tab1}.
It can be seen that our proposed UMM-CSGM method achieved the best synthesis performance for all tested missing-modality configurations, in terms of the average values of all three evaluation metrics. Meanwhile, the performance variation is lower using our methods especially when synthesizing Flair and T1 images. 
Therefore, our method \emph{not only} yields better synthesis accuracy, \emph{but also} can be more robust to address some hard cases. The performance gain achieved by our UMM-CSGM method indicates that it can learn better representation of the cross-modal relationships.

\subsubsection{Qualitative results.}
For visual evaluation, Fig.~\ref{fig2} demonstrates two sets of representative synthesized images of various missing modalities using different conditional generative methods. The left and right panel show results with benign and malignant tumors, respectively.
It can be seen that UMM-CSGM can provide more structural details and clearer textural features in the brain tissue, which can be more prominently appreciated in the T1 images of benign tumor (the second column in Fig.~\ref{fig2}). It places a much clearer description of the gray and white matters as notified in the red box, i.e., the gray and white matter are more distinctive. 
More importantly, our method performs better in the lesion area affected by malignant tumors. Particularly, the tumor-affected areas can sometimes present themselves as brighter contrast and sometimes darker contrast in Flair images (the seventh column in Fig.~\ref{fig2}). Existing methods have problem handling such diversity, leading to artificial signal enhancement in the lesion. In contrast, our method can perform significantly better, without implementing spatial attentions, extracting complex textural features, or demanding more conditional information compared to competing methods. The better cross-modal dependence modeling ability of UMM-CSGM itself can automatically address the requirement of diversity.

\section{Conclusion}
In this paper, to address the deficiency of deterministic mapping in modeling cross-modal relationships for multi-modal medical image completion, we propose a Unified Multi-Modal Conditional Score-based Generative Model (UMM-CSGM) framework. 
This probabilistic framework uses only a single score-based network to learn various cross-modal conditional distributions in the complete modality space, and can effectively and robustly infer any missing modality based on the learned conditional distribution by leveraging all remaining modalities for accurate conditioning. Different from the existing GAN-based mapping methods, the uncertainties in the cross-modal relationships are incorporated by UMM-CSGM into a stochastic iterative refinement process that gradually refines the pure noise to the missing modality under the guidance of the available modality. Compared to the SOTA methods, extensive experiments on BraTS19 dataset demonstrate that UMM-CSGM can better capture and more effectively sample the cross-modal relationships to infer the missing modality, which is reflected as more discriminative delineation of the complex structures and textures (i.e., higher fidelity), and more reliable inference of the heterogeneous contrast within malignant tumors in synthesized brain MRI (i.e., higher diversity).

\bibliographystyle{splncs04}
\bibliography{ref.bib}
\end{document}